\documentclass[preprint,epsfig,onecolumn,floats,showpacs]{revtex4}
\usepackage{amssymb}
\usepackage{graphicx}
\usepackage{epsfig}
\usepackage{amsmath}
\usepackage{amsfonts}
\usepackage{bm}
\usepackage{subfigure}

\begin{document}

\title{Substrate Modulated Graphene Quantum Dot}
\author{Qiong Ma}
\author{Zhi-Rong Lin}
\author{Tao Tu}
\email{tutao@ustc.edu.cn}
\author{Guang-Can Guo}
\author{Guo-Ping Guo}
\email{gpguo@ustc.edu.cn}
\affiliation{Key Laboratory of Quantum Information, University of Science and Technology
of China, Chinese Academy of Sciences, Hefei, 230026, P.R.China}
\date{\today}

\begin{abstract}
We propose a new method to use gapped graphene as barrier to confine
electrons in gapless graphene and form a good quantum dot, which can be
realized on an oxygen-terminated $SiO_{2}$ substrate partly H-passivated. In
particular, we use ferromagnetic insulators deposited on top of barrier
which give rise to a spin related energy spectrum and transport properties.
Compared to the complexity of etched quantum dots in graphene, the setup
suggested here is a promising candidate for practical applications.
\end{abstract}

\maketitle

\baselineskip 16pt

\textit{Introduction.} Graphene has attracted a lot of research interest
because of its unique electronic properties which make it a promising
candidate for future nanoelectronics \cite%
{Geim2004,Geim2005,Kim2005,Heer2006,Dai2008}. However, there are still many
barriers in the way of making effective uses out of it. For example, due to
the absence of gap between conductance and valance bands in the carrier
spectrum and the phenomenon of Klein tunnelling, it is hard to confine
electrons within a small region to form quantum dot using electrostatic
potential barriers \cite{Geim2006}. Alternative strategies have been
proposed to solve this difficulty by etching graphene into nanostructures
\cite{Geim2008,Ensslin2008}, using non-zero transverse momentum in armchair
nanoribbon \cite{Efetov2007,Burkard2007}, considering bilayer structure \cite%
{Peeter2007}, or applying inhomogeneous magnetic fields \cite{Egger2007}.
Here, we propose a new and easier method to use gapped graphene as barrier
to confine electrons in gapless graphene and form a good quantum dot.

\textit{Setup. }It is well known that electron in ideal graphene behaves as
a massless Dirac-fermion whose energy spectrum has no gap between conduction
and valance bands. Recently, the electron in epitaxially grown graphene
monolayer on a $SiC$ substrate is found to be massive close to the Dirac
point because of the symmetry breaking of sublattice caused by substrate and
lattice interaction \cite{SiC2007}. Further, there are also some experiments
carried on widely used $SiO_{2}$ substrate \cite{Ishigami2007}. Ref. \cite%
{Philip2009} points out that if a single layer graphene is deposited onto a $%
SiO_{2}$ surface, the electronic energy spectrum of the monolayer graphene
depends strongly on the surface characteristic, i.e. a finite energy gap
will open between conduction and valence bands for an oxygen-terminated
surface, but close when the oxygen atoms on the substrate are passivated
with hydrogen atoms. Thus if an oxygen terminated $SiO_{2}$ substrate is
fully exposed to hydrogen atoms atmosphere within a small region in the
middle, then a single layer graphene is deposited on it, and we will get a
gapless part confined by gapped parts, which can serve as barriers, as shown
in Fig. 1. Moreover, when we make the barrier region also exposed to
hydrogen atoms atmosphere but in a different degree from the dot, the gap
will not close completely, and we can tune the barrier height in this
system. It is realizable in experiment by using PMMA to cover the region
which doesn't need to be passivated.

\begin{figure}[tbp]
\centering
\includegraphics[width=0.7\columnwidth]{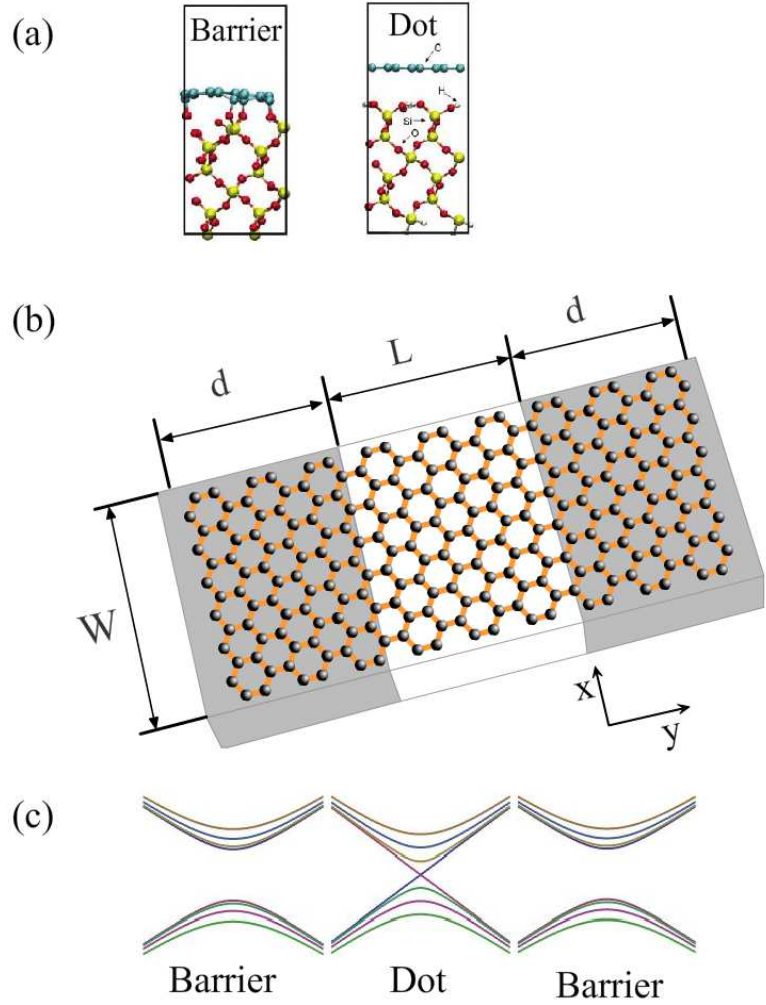}
\caption{ Schematic illustration of a graphene quantum dot. (a) The
substrate and graphene interaction when hydrogen-passivated (right) and non
hydrogen passivated (left). (b) The light part is the dot region, which is
fully hydrogen-passivated and gapless. The dark part is the barrier region,
which is non hydrogen-passivated or slightly hydrogen-passivated and gapped.
(c) The energy spectrum of this system.}
\end{figure}

For simplicity and clarity, we put our discussion in the setup of metallic
armchair shaped graphene nanoribbon. In the present case, we use the
substrate induced energy gap as barrier to confine electrons, therefore the
realization of quantum dot will not depend much on the boundary conditions.
More remarkably, we consider adding ferromagnetic insulator such as $EuO$
upon the two gapped graphene barriers. Ferromagnetic insulators deposited on
graphene can induce ferromagnetic correlations in graphene and the induced
exchange interaction is estimated to achieve $5$ meV by using $EuO$ \cite%
{Haugen2008}. We find that it will lead to spin dependent energy spectrum
and conductance phenomenon in the proposed graphene dot.

\textit{Bound States and Energy Spectrum.} The electron waves in graphene
system are usually described by four component spinor envelop wavefunction $%
\Psi =(\psi _{A}^{(K)},\psi _{B}^{(K)},-\psi _{A}^{(K^{^{\prime }})},-\psi
_{B}^{(K^{^{\prime }})})$ and their behaviors will be governed by $4\times 4$
Dirac equation for massless or massive particles, which can be written as:
in the dot region (where $0\leq y\leq L$),
\begin{equation}
-i\hbar v_{F}\left(
\begin{array}{cc}
{\sigma _{x}\partial _{x}+\sigma _{y}\partial _{y}} & 0 \\
0 & {-\sigma _{x}\partial _{x}+\sigma _{y}\partial _{y}}%
\end{array}%
\right) \Psi =E\Psi ,
\end{equation}%
and in the barrier region (where $y<0$ or $y>L$),
\begin{equation}
-i\hbar v_{F}\left(
\begin{array}{cc}
{\sigma _{x}\partial _{x}+\sigma _{y}\partial _{y}} & 0 \\
0 & {-\sigma _{x}\partial _{x}+\sigma _{y}\partial _{y}}%
\end{array}%
\right) \Psi +\Delta \left(
\begin{array}{cc}
\sigma _{z} & 0 \\
0 & \sigma _{z}%
\end{array}%
\right) \Psi -\eta V_{\sigma }\Psi =E\Psi ,
\end{equation}%
where $\hbar $ is the Planck constant divided by $2\pi $, $v_{F}\approx
10^{6}$ m/s is the analog of the Dirac electron speed of light, $\sigma _{x}$%
, $\sigma _{y}$, $\sigma _{z}$ are Pauli matrices acting on two-spinor
states related to the two triangular sublattices of graphene, $\eta =\pm 1$
stands for the two spin indexes. $2\Delta $ is the induced gap by the
substrate, $2V_{\sigma }$ is the spin splitting energy due to the
correlation with ferromagnetic contacts.

We consider metallic armchair boundaries and the quantized transverse
momentum $q_{n}=n\frac{\pi }{W}$ keeps the same during the movement of
electron, but the wave vector in the $y$ direction must satisfy different
conditions as
\begin{equation}
E=\pm \sqrt{(\hbar v_{F}q_{n})^{2}+(\hbar v_{F}k)^{2}},
\end{equation}%
in the dot and%
\begin{equation}
E=\pm \sqrt{(\hbar v_{F}q_{n})^{2}+(\hbar v_{F}k^{\prime })^{2}+\Delta ^{2}}%
-\eta V_{\sigma },
\end{equation}%
in the barriers, where $k$ is the wave vector in the dot and $k^{^{\prime }}$
in the ferromagnetic barrier with $\pm $ signs referring to conduction band $%
(+)$ and valence band $(-)$ respectively. The bound state requires that $%
k^{^{\prime }}$ is a pure imaginary, which means the bound state energy
should satisfy
\begin{equation}
|E|\geqslant \hbar v_{F}|q_{n}|,\text{ }|E+\eta V_{\sigma }|<\sqrt{(\hbar
v_{F}q_{n})^{2}+\Delta ^{2}}.
\end{equation}

\begin{figure}[tbp]
\centering
\subfigure[]{\label{2a}\includegraphics[width=0.7%
\columnwidth]{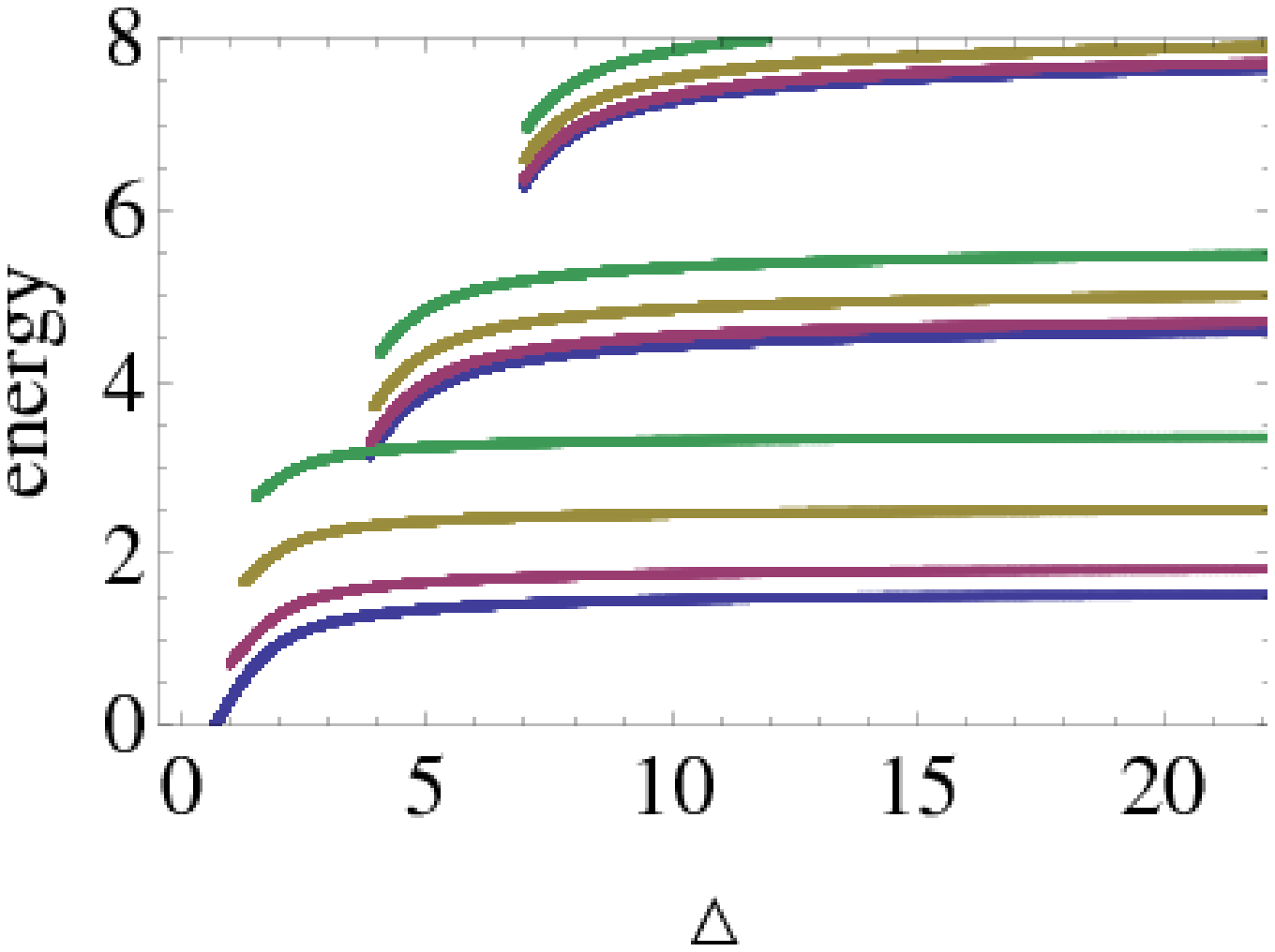}} \subfigure[]{\label{2b}%
\includegraphics[width=0.35\columnwidth]{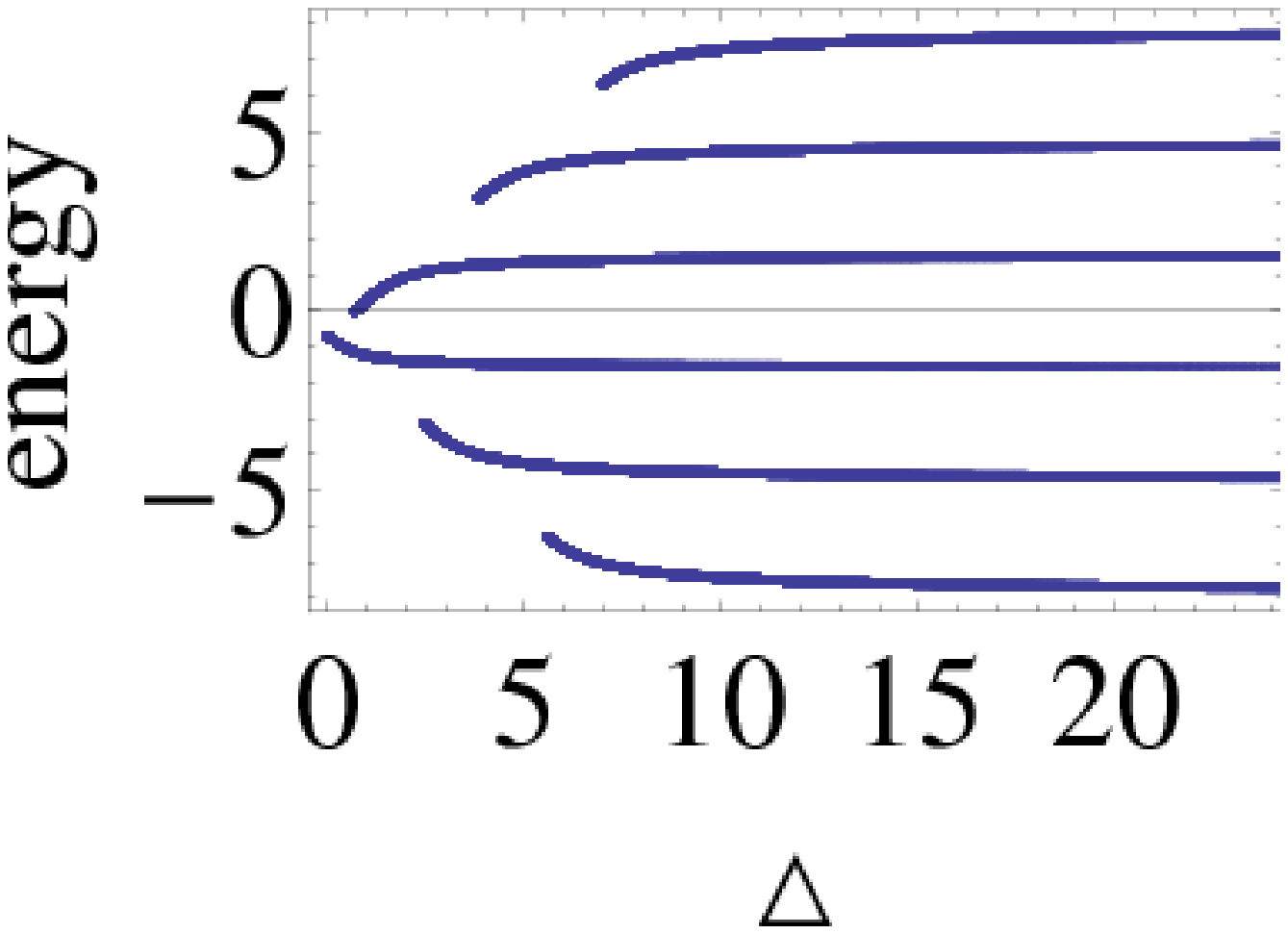}} %
\subfigure[]{\label{2c}\includegraphics[width=0.35%
\columnwidth]{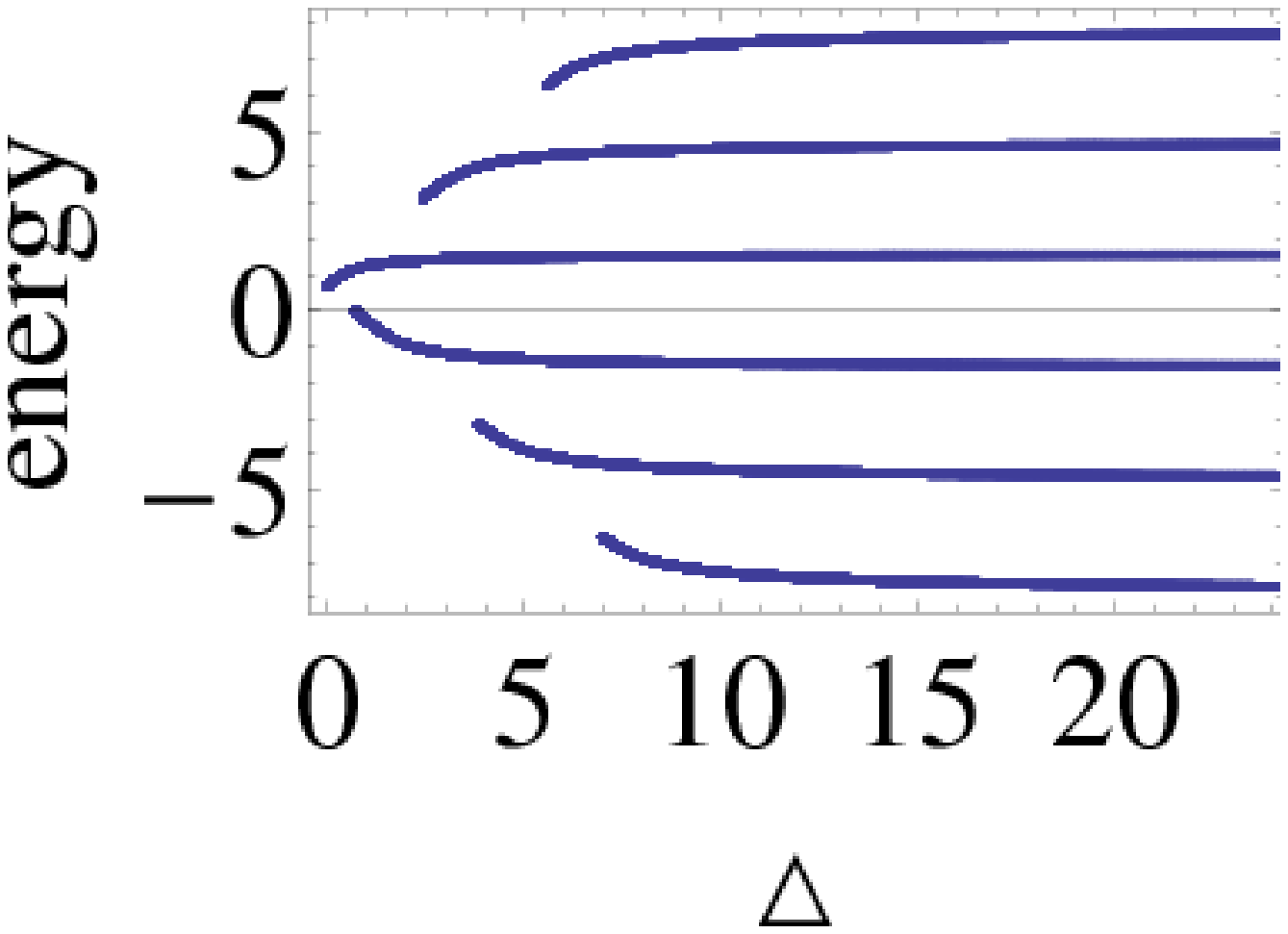}} \subfigure[]{\label{2d}%
\includegraphics[width=0.35\columnwidth]{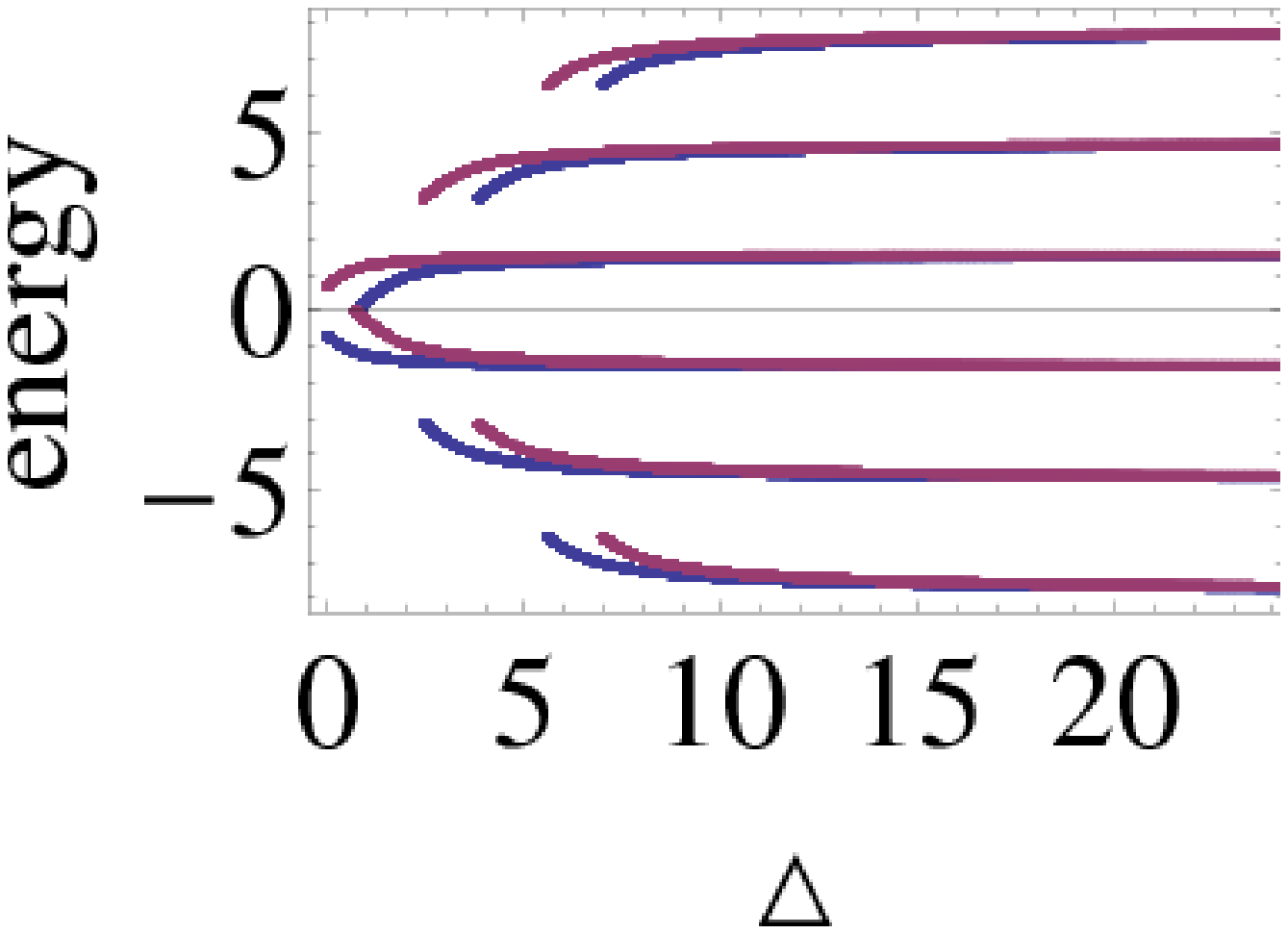}} %
\subfigure[]{\label{2e}\includegraphics[width=0.35%
\columnwidth]{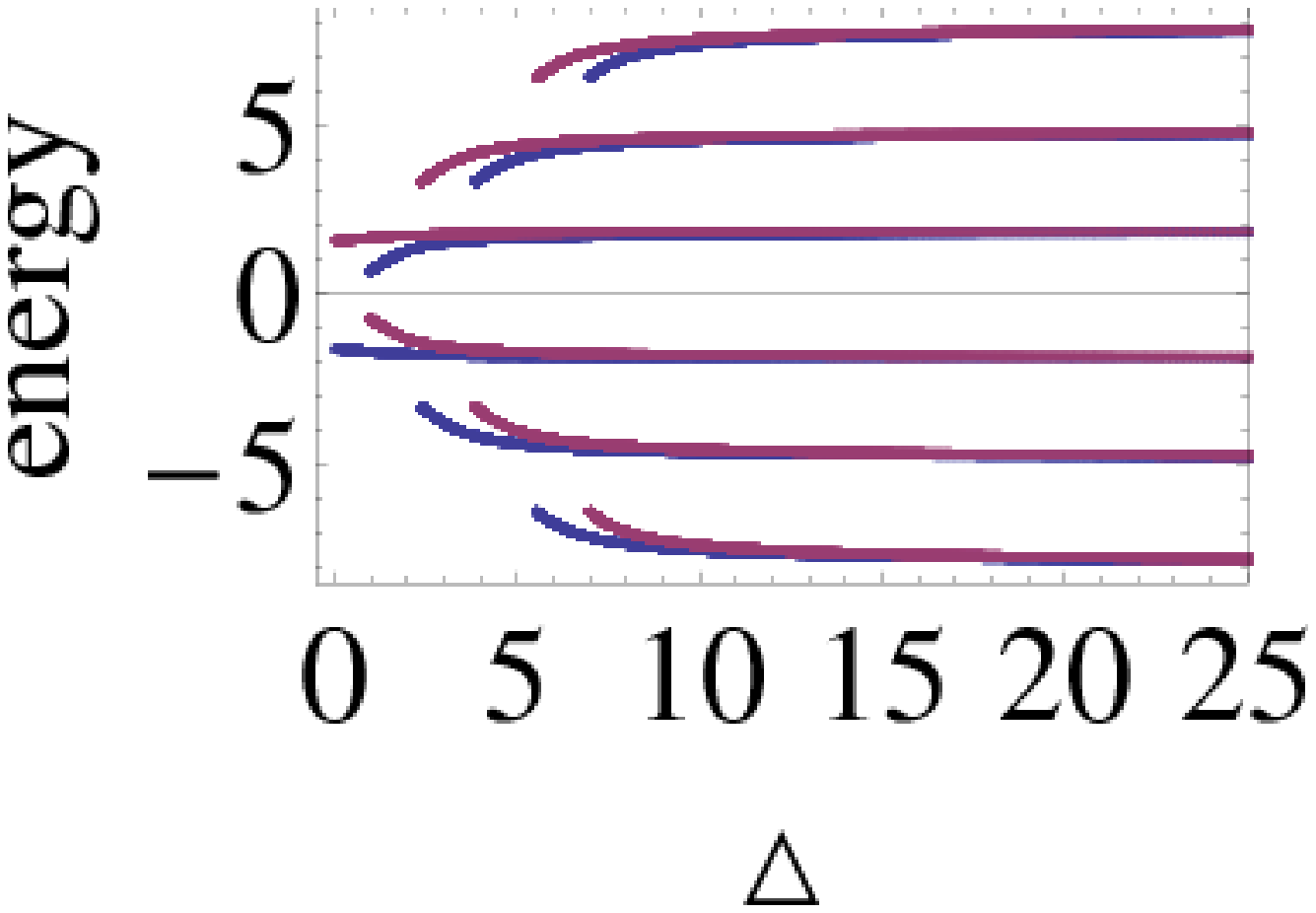}}
\caption{ Bound-state energy levels in the assumption $V_{\protect\sigma}=5
meV$. Both axe labels are in the characteristic energy unit of $\hbar
v_{F}/L $. L is the length of dot, and if L equals 100 nm, then the
characteristic energy unit is about 7 meV. (a) for $q_{n}=0,1,2,3$ and $%
\protect\eta=1$, blue: $q_{n}=0$, red: $q_{n}=1$, yellow: $q_{n}=2$, green: $%
q_{n}=3$. (b) $q_{n}=0$ and $\protect\eta=1$, (c) $q_{n}=0$ and $\protect\eta%
=-1$, (d) $q_{n}=0$, the blue line and red line respectively stand for $%
\protect\eta=1$ and $\protect\eta=-1$, (e) $q_{n}=1$, the blue line and red
line respectively stand for $\protect\eta=1$ and $\protect\eta=-1$. }
\end{figure}

After matching the wavefunctions in different regions at $y=0$ and $y=L$, we
can get the energy levels of the bound states. We set the parameters of this
system as $L=100$ nm and $W=\pi L\approx 300$ nm. Then if we use $1/L$ as
the unit of $q_{n}$, $q_{n}=n$. The characteristic energy $\hbar v_{F}/L$,
which is about $7$ meV in this case, will be used as the energy unit below.
In Fig. 2, we show the energy spectrum as a function of the substrate
induced interaction $\Delta $ for different transverse momentums ($q_{n}$)
and spin indexes ($\eta $) where $V_{\sigma }$ is assumed to be $5$ meV
according to Ref. \cite{Haugen2008}. Fig. 2a plots the energy spectrum for $%
q_{n}=0,1,2,3$ and $\eta =1$ above the Dirac point. When $\Delta $
increases, the number of bound states is increasing at the same time, which
can be deduced from Eq. 5. Bound states are formed even when the transverse
momentum is zero, which is distinguished from the former results \cite%
{Burkard2007}. The result lies at the heart of our approach where the dot
levels (bound states) are located in the gap of the barrier regions induced
by the interaction with substrate, which is schematically illustrated in the
Fig. 1c. More interestingly, we find that for a particular spin index, the
energy spectrum is unsymmetrical relative to the zero energy point, as shown
in Fig. 2b for $\eta =1$ and Fig. 2c for $\eta =-1$. Non-zero $V_{\sigma }$
shift the potential for a certain spin orientation only in the barrier
region, therefore the symmetry of the spectrum in the dot region shown in
Fig. 1c are broken. However the chirality between electron and hole remains
if the two spin indexes are considered together, as shown in Fig. 2d and
Fig. 2e. What should be emphasized is that symmetric states here belong to
opposite spin indexes. Besides, we can find that with the increase of $%
\Delta $, the difference between energy of different spins is suppressed,
and each energy level becomes spin degenerate again.

\begin{figure}[tbp]
\centering
\subfigure[]{\label{3a}\includegraphics[width=0.5\columnwidth]{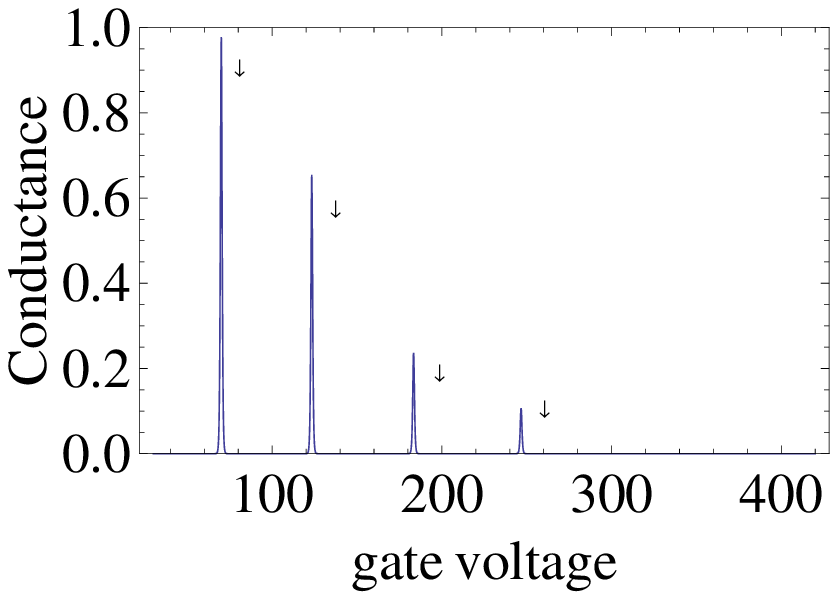}}
\subfigure[]{\label{3b}\includegraphics[width=0.5\columnwidth]{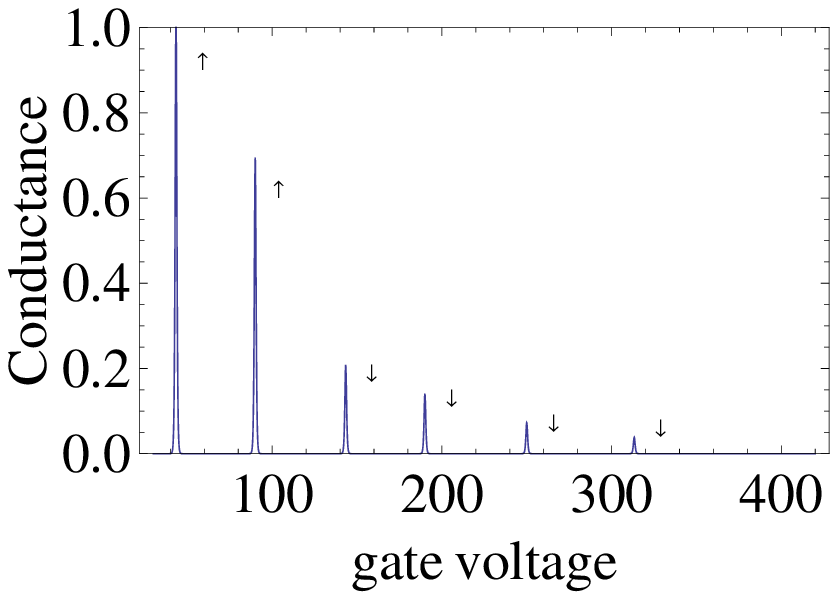}}
\subfigure[]{\label{3c}\includegraphics[width=0.5\columnwidth]{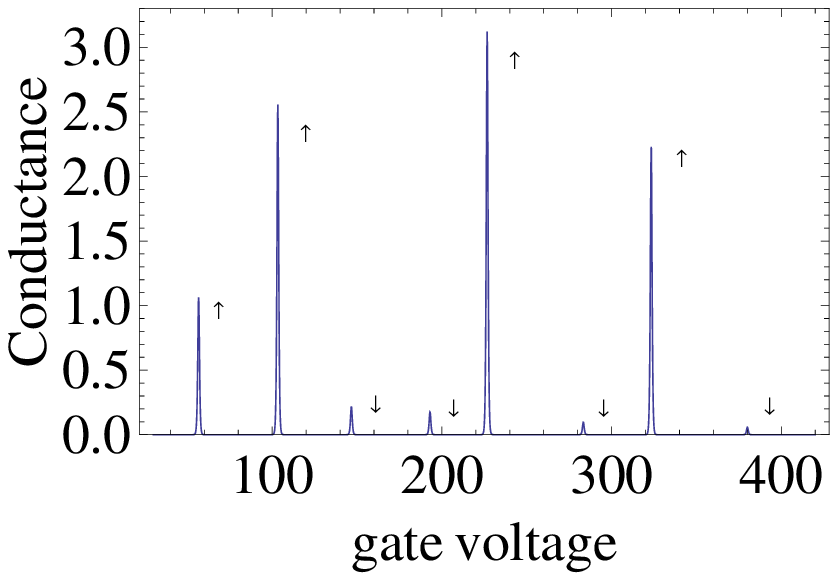}}
\caption{ (a) Coulomb Blockade oscillation at different $\Delta $ values
when $V_{\protect\sigma }=0.7$ and $k_{B}T=0.01$. The gate voltage is in a
unit of meV and the conductance is in a unit of $\frac{e^{2}}{4k_{B}T}\Gamma
_{1}$. (a) $\Delta =0.5$, (b) $\Delta =1$, (c) $\Delta =1.5$. The up arrow
stands for $\protect\eta =1$, and the down arrow stands for $\protect\eta %
=-1 $. $\Delta $, kT, and $V_{\protect\sigma }$ are all in a unit of $\hbar
v_{F}/L$.}
\end{figure}

\textit{Coulomb Blockade Behaviors.} In the following part, we will study
the transport properties of the system above \cite{Kouwenhoven2001}. If a
bias voltage $V_{sd}$ is applied between left and right reservoirs, a
current $I$ can pass through the dot. The number of electrons in the dot,
and hence its energy is varied by an applied gate voltage. In the liner
conductance response regime ($V_{sd}\approx 0$), we will observe the Coulomb
Blockade of single tunneling process, which will lead to a series of sharp
peaks, as long as heat fluctuation cannot compete with the energy separation
\cite{Beenakker1990}. First, we adopt the constant interaction model, which
assumes that the Coulomb interaction between the electrons is independent of
the number $N$ and can be described by a constant capacitance $C$, and
estimate the charging energy $e^{2}/C\approx 1$ in a unit of $\hbar v_{F}/L$
, where $e$ is the charge of the electron. Then we use the method described
in Ref. \cite{Beenakker1990} to discuss in detail this single electron
tunneling phenomenon at low temperature. The linear response conductance is
\begin{equation}
G=-\frac{e^{2}}{2k_{B}T}\underset{N}{\sum }\Gamma _{N}\ f^{\prime
}(E_{N}+U(N)-U(N-1)-E_{F})
\end{equation}%
where $f(x)=\frac{1}{1+e^{\frac{x}{k_{B}T}}}$ is the Fermi-Dirac
distribution function, $E_{N}$ is the energy of the top filled single
electron state for a $N$ electron dot, and $U(N)=(Ne^{2})/2C-Ne\phi _{ext}$,
in which $\phi _{ext}=\phi _{0}+\alpha V_{gate}$. The width of localized
energy $\Gamma $ in the above equation is determined by tunneling though the
classically forbidden region \cite{Chakraborty2007},
\begin{equation}
T=exp\left( -\int_{L}^{L+d}\left\vert k^{^{\prime }}\right\vert dy\right)
\end{equation}%
where $k^{\prime }=i\sqrt{q_{n}{}^{2}+(\frac{\Delta }{\hbar v_{F}})^{2}-(%
\frac{E+\eta V_{\sigma }}{\hbar v_{F}})^{2}}$ is the vanishing wave vector
in the barrier region and $d$ is the width of each barrier. Here, we assume $%
d=L$. It must be emphasized that the tunneling rate is not precisely gained
theoretically, so the amplitude of Coulomb Blockade peaks can only show a
general pattern. Inversely, the tunneling rate can be measured from
experiment by studying the amplitude of Coulomb Blockade peaks. Fig. 3 shows
the obtained conductance as a function of gate voltage at different $\Delta $
values assuming that energy levels below the Dirac point have already been
filled up. Here, the up arrow stands for $\eta =1$ and the down arrow stands
for $\eta =-1$. From Fig. 3, we can see that when $\Delta $ is very small,
all the conductance peaks for up spin have been suppressed, only down spin
peaks remain. When $\Delta $ increases, peaks for up spin appear. This
simply derives from the fact that for $\Delta <V_{\sigma }$, there is no
bound states for spin up case and for $\Delta >V_{\sigma }$, the spin up
energy level appears later than spin down, as shown in Fig. 2d and Fig. 2e.
Here, the spin down or up is relative, and they can be exchanged if we
change the direction in which the ferromagnetic insulator is deposited.

\textit{Conclusion.} In this paper, we propose a new method to form a
quantum dot in graphene without electrostatic barrier. By using
oxygen-terminated $SiO_{2}$ substrate which is partly passivated by Hydrogen
atom, we can realize to make gapped graphene around gapless graphene. The
gapped graphene serves as a natural barrier for gapless graphene due to its
substrate induced opening energy gap. In particular, we use ferromagnetic
insulators deposited on top of gapped graphene to induce a energy splitting
between spin up and down levels. We systematically investigate the bound
states of the dot and get the energy spectrum for different spins as a
function of substrate induced energy gap. We also study the transport
behavior in the system and show how the liner response conductance for
different spins is modified by the change of substrate induced energy gap.
Compared to the complexity of experiments on etched quantum dots in graphene
\cite{Geim2008,Ensslin2008}, the setup suggested here has potential to
become a well tunable nanodeivice using today's fabrication techniques, and
can be directly developed to array of many quantum dots. This unique feature
is of practical importance for future applications in quantum computations
\cite{Burkard2007}.

\textbf{Acknowledgement} This work at USTC was funded by National Basic
Research Programme of China (Grants No. 2006CB921900 and No. 2009CB929600),
the Innovation funds from Chinese Academy of Sciences, and National Natural
Science Foundation of China (Grants No. 10604052 and No. 10874163 and
No.10804104).


\end{document}